\documentclass[11pt]{article}
\newsavebox{\PSLASH}
\sbox{\PSLASH}{$p$\hspace{-1.8mm}/}

\input{amssym}
\def\be{\begin{equation}}
\def\ee{\end{equation}}
\def\ba{\begin{eqnarray}}
\def\ea{\end{eqnarray}}
\begin{document}
\title{\large \bf Thick planar domain wall: its thin wall limit and dynamics}
\author{S. Ghassemi$^{1}$\footnote{e-mail:
ghassemi@ipm.ir}, S. Khakshournia$^{2}$\footnote{e-mail:
skhakshour@aeoi.org.ir}, and R. Mansouri$^{3}$\footnote{e-mail:
mansouri@ipm.ir}  \\ \\
$^{1,3}$Institute for Studies in Physics and Mathematics (IPM),
Tehran, Iran\\
$^{2}$Nuclear Research Center, Atomic Energy Organization of Iran,
Tehran, Iran\\
$^{3}$Department of Physics, Sharif University of
Technology, Tehran, Iran\\
}\maketitle
\[\]
\[ \]
\begin{abstract}
We consider a planar gravitating thick domain wall of the $\lambda
\phi^4$ theory as a spacetime with finite thickness glued to two
vacuum spacetimes on each side of it. Darmois junction conditions
written on the boundaries of the thick wall with the embedding
spacetimes reproduce the Israel junction condition across the wall
in the limit of infinitesimal thickness. The thick planar domain
wall located at a fixed position is then transformed to a new
coordinate system in which its dynamics can be formulated. It is
shown that the wall's core expands as if it were a thin wall. The
thickness in the new coordinates is not constant anymore and its
time dependence is given.
\end{abstract}
\newpage
\section{Introduction}
Domain walls are solutions to the coupled Einstein-scalar field
equations with a potential having a spontaneously broken discrete
symmetry and a discrete set of degenerate minima. In the simplest
case of two minima, a domain wall having a non-vanishing energy
density appears in the separation layer, with the scalar field
interpolating between these two values.\\
Domain walls in the cosmological context have a long history
\cite{vilen}. It was realized very early that the formation of
domain walls with a typical energy scale of $\geq 1 MeV$ must be
ruled out \cite{zel}, because a network of such objects would
dominate the energy of the universe, violating the observed
isotropy and homogeneity. Domain walls were reconsidered in a
possible late time phase transition scenario at the scale of $\leq
100 MeV$. Such walls were supposed to be thick because of the low
temperature of the phase transition \cite{hill}. The suggestion
that Planck size topological defects could be regarded as triggers
of inflation, revived the discussion of thin and thick domain
walls \cite{vil2}. The realization of our universe as a
$(3+1)$-dimensional domain wall immersed in a higher dimensional
spacetime has led to the recent numerous studies \cite{rand}.\\
First attempts to investigate the gravitational properties of
domain walls were based on the so called thin wall approximation.
In this approach one forgets about the underlying field theory and
simply treats the domain wall as a zero thickness
(2+1)-dimensional timelike hypersurface embedded in a
four-dimensional spacetime. The Darmois-Israel thin wall formalism
\cite{israel} is then used to match the solutions of the Einstein
equations on both sides of the wall in the embedding spacetime
across the thin wall. However, such spacetimes have delta
function-like distributional curvature and energy-momentum tensor
supported on the hypersurface. Using this procedure, the first
vacuum solutions for a spacetime containing an infinitely thin
planar domain wall was found by Vilenkin \cite{vil3}, and Ipser
and Sikivie \cite{ipser}. The very interesting feature of such
domain walls is that they are not static, but have a de
Sitter-like expansion in the wall's plane. External observers
experience a repulsion from the wall, and there is an event
horizon at finite proper distance from the wall's core. These
results were initially obtained within the framework of the
Darmois-Israel thin wall formalism which has been used as an
approximative description of real wall of finite thickness.
Typically, a self gravitating domain wall has two length scales,
its thickness $w$ and its distance to the event horizon which can
be compared to $w$. Since these lengths are expressed in terms of
the coupling constants of the theory, thin walls turn out to be an
artificial construction in terms of these underlying parameters
as mentioned in \cite{BCG1}.\\
The first exact dynamical solution to thick planar domain walls
was obtained by Goetz \cite{goetz} and later a static solution was
recovered with the price of sacrificing reflection symmetry
\cite{mukh}. Silveira \cite{silv} studied the dynamics of a
spherical thick domain wall by appropriately defining an average
radius $<R>$, and then used the well-known plane wall scalar field
solution as the first approximation to derive a formula relating
$<\ddot{R}>$, $<\dot{R}>$, and $<R>$ as the equation of motion for
the thick wall. Widrow \cite{wid} used the Einstein-scalar
equations for a static thick domain wall with planar symmetry. He
then took the zero-thickness limit of his solution and showed that
the orthogonal components of the energy-momentum tensor vanish in
that limit. Garfinkle and Gregory \cite{gar} presented a
modification of the Israel thin shell equations to treat the
evolution of thick domain walls in vacuum. They used an expansion
of the coupled Einstein-scalar field equations describing the
thick gravitating wall in powers of the thickness of the domain
wall around the well-known solution of the hyperbolic tangent kink
for a $\lambda\phi^{4}$ wall and concluded that the effect of the
thickness at the first approximation was to effectively reduce the
energy density of the wall compared to the thin case, leading to a
faster collapse of a spherical wall as well as a slower expansion
of a planer wall in vacuum. Barrab\'{e}s et al. \cite{bar} applied
the expansion in the wall action and integrated it out
perpendicular to the wall to show that the effective action for a
thick domain wall in vacuum, apart from the usual Nambu term,
consists of a contribution proportional to the induced Ricci
curvature scalar. Arod\'{z} et al. \cite{Arodz} applied a cubic
polynomial ansatz for the scalar field inside the wall while the
field gets the vacuum values outside the wall. Then the field
equations and boundary conditions were used to determine the
coefficients of polynomial. They obtained a Nambu-Goto type for
the core of the wall together with a nonlinear equation for its
thickness. Within the context of a fully nonlinear treatment of a
scalar field coupled to gravity, Bonjour, Charmousis and Gregory
(BCG) found an approximate but analytic description of the
spacetime of a thick planar domain wall of the $\lambda \phi^4$
model by examining the field equations perturbed in a parameter
characterizing the gravitation interaction of the scalar field
\cite{gre99}, but they did not analyze the motion of the thick
wall. The thin wall limit of Goetz's solution was studied in
\cite{rom} and it was shown that this solution has a well-defined
limit. \\
A completely different method based on the gluing of a thick wall,
considered as a regular manifold, to two different manifolds on
each side of it was developed by Khakshournia-Mansouri (KM) in
\cite{khak02} and recently in a more generalized context in
\cite{khos}. In this paper, we will first investigate the thin
wall limit of the thick BCG-like domain wall adapted to the KM
formalism. We then introduce a coordinate transformation from the
BCG solution to the one in which one can observe evolution of the
thick wall.\\
The organization of this paper is as follows. In section 2 we give
a brief introduction to BCG thick wall solution and summarize all
the useful equations we need in this paper. Section 3 is devoted
to the thick wall formalism developed in \cite{khak02} and its
application to the thick planar domain wall. The thin wall limit
is then established in section 4. The dynamics of this thick
domain wall in the new coordinates is studied in section 5. We
then summarize our results in section 6.
\section{Planar domain wall solution}
Domain walls as regions of varying scalar field separating two
vacua with different field values are usually described by the
following matter Lagrangian:
\begin{equation}\label{Lagrangian}
L=\nabla_{\mu}\phi\nabla^{\mu}\phi-V(\phi),
\end{equation}
where $\phi$ is a real scalar field and $V(\phi)$ is a symmetry
breaking potential which we take it to be
$V(\phi)=\lambda(\phi^2-\eta^2)^2$, where $\lambda$ is a coupling
constant and $\eta$ is the symmetry breaking scale. Looking for a
static solution of the equation of motion derived from this
Lagrangian in flat space-time, one gets
\begin{equation}\label{phi}
X=\tanh\frac z w,
\end{equation}
where $X=\frac\phi\eta$, $w=\frac{1}{\sqrt{\lambda}\eta}$, and $z$
is the coordinate normal to the wall. This particular solution
represents an infinite planar domain wall centered at $z=0$. From
the stress-energy tensor of the wall, one can easily observe that
the wall energy density is peaked around $z=0$ and falls down
effectively at $z=w$. We may therefore call $w$, being a
characteristic length scale of the domain wall, the effective
thickness of the wall.\\
We now look at the planar gravitating domain wall solutions. The
line element of a plane symmetric spacetime may be written in the
general form
\begin{equation}\label{nmet}
ds^2=A^2(z)dt^2-B^2(z,t)(dx^2+dy^2)-dz^2,
\end{equation}
which displays reflection symmetry around the wall's core assuming
to be located at z=0, where z is the proper length along the
geodesics orthogonal to the wall. In order to obtain a thick
gravitating wall solution one should solve the coupled system of
the Einstein and scalar field equations which are as follows
\begin{equation}\label{Richi}
R_{\mu\nu}=\epsilon\left(2X_{,\mu}X_{,\nu}-\frac{1}{w^2}
g_{\mu\nu}(X^2-1)^2\right),
\end{equation}
\begin{equation}\label{field}
\square X+\frac{2}{w^2}X(X^2-1)=0,
\end{equation}
where $R_{\mu\nu}$ is the Ricci tensor. For a static field $X(z)$,
the Einstein equations (\ref{Richi}) leads to the constraint
$B(z,t)=A(z)\exp(kt)$.\\
BCG investigated the spacetime of a thick gravitating planar
domain wall for a $\lambda \phi^4$ potential \cite{gre99}. In this
context, the dimensionless parameter $\epsilon$ in front of the
equation (\ref{Richi}) can be taken to characterize the coupling
of gravity to the scalar field:
\begin{equation}\label{epsilon}
\epsilon\equiv 8\pi G\eta^2.
\end{equation}
Supposing that gravity is weakly coupled to the scalar field,
$A(z)$ and $X(z)$ may be expanded in the powers of $\epsilon$:
\begin{eqnarray}
A_i(z)=A_0(z)+\epsilon A_1(z)+O(\epsilon^2),\\\
X(z)=X_0(z)+\epsilon X_1(z)+O(\epsilon^2),
\end{eqnarray}
where we have introduced the index $i$ to indicate the quantities
inside the wall.\\
In the limit $\epsilon\rightarrow  0$, one should obtain the same
results as that of the non-gravitating planar wall, i.e.
$A_i(z)=1$ and $X(z)=\tanh(\frac z w)$. Using these expansions,
BCG solved the coupled Einstein and scalar field equations to
first order in $\epsilon$ and obtained the following results:
\begin{equation}\label{ain}
A_i(z)=1-\frac{\epsilon}{3}\left[2\ln(\cosh\frac z w)+\frac 1 2
\tanh^2\frac z w\right]+O(\epsilon^2),
\end{equation}
\begin{equation}\label{kin}
k=\frac 2 3\frac{\epsilon}{w}+O(\epsilon^2),
\end{equation}
\begin{equation}
\label{inside metric3} X(z)=\tanh\frac z w-\frac{\epsilon}{2}{\rm
sech}^2\frac z w \left(\frac z w+\frac 1 3 \tanh\frac z w \right)+
O(\epsilon^2).
\end{equation}
Eqs. (\ref{ain}-\ref{inside metric3}) represent the perturbative
solution to the spacetime of the thick wall described by Eq.
(\ref{nmet}). In the presence of gravity, there is an event
horizon at the proper distance $z=\frac{1}{k}$ from the core of
the wall where $A_i(z)$ goes to zero and the coordinate system
used in (\ref{nmet}) breaks down as discussed by BCG.
\section{The thick wall formalism}
In this section we first make a short review of the thick wall
formalism developed by  KM in \cite{khak02}. Then we apply it to
the thick planar domain wall described by the metric (\ref{nmet}).
In KM formalism a thick wall is modelled with two boundaries
$\Sigma_1$ and $\Sigma_2$ dividing a spacetime $\cal M$ into three
regions. Two regions $\cal M_{+}$ and $\cal M_{-}$ on either side
of the wall and region $\cal M$$_{0}$ within the wall itself.
Treating the two surface boundaries $\Sigma_1$ and $\Sigma_2$
separating the manifold $\cal M$$_{0}$ from two distinct manifolds
$\cal M_{+}$ and $\cal M_{-}$, respectively, as nonsingular
timelike hypersurfaces, we do expect the intrinsic metric
$h_{\mu\nu}$ and extrinsic curvature tensor $K_{\mu\nu}$ of
$\Sigma_j\hspace{0.2cm}$ (j=1,2) to be continuous across the
corresponding hypersurfaces. These requirements named the Darmois
conditions are formulated as
\begin{equation}\label{hmn}
[h_{ab}]_{\Sigma_j}=0\hspace {1cm} j=1,2,
\end{equation}
\begin{equation}\label{thick israel}
[K_{ab}]_{\Sigma_j}=0\hspace {1cm} j=1,2,
\end{equation}
where the square bracket denotes the jump of any quantity that is
discontinuous across $\Sigma_j$, Latin indices range over the
intrinsic coordinates of $\Sigma_{j}$, and Greek indices over the
coordinates of the 4-manifolds. To apply the Darmois conditions on
two surface boundaries of a given thick wall one needs to know the
metrics in three distinct spacetimes $\cal M_{+}$, $\cal M_{-}$
and $\cal M$$_{0}$ being jointed at $\Sigma_j$. While the metrics
in $\cal M_{+}$ and $\cal M_{-}$ are usually given in advance,
knowing the metric whin the wall spacetime $\cal M$$_{0}$
requires a nontrivial work. \\
Let us now impose these junction conditions for the self
gravitating thick planar domain wall described in the previous
section. Recalling $\epsilon\ll 1$, corresponding to weak self
gravity, we first introduce a dimensionless parameter
$\frac{1}{\epsilon}>\Delta \gg1$ to assume the scalar field takes
its vacuum values on the wall boundaries $\Sigma_1$ and $\Sigma_2$
being located at the proper distances $z=\pm \:\: \Delta w/2$ far
from the wall's core surface  while the coordinate system in
(\ref{nmet}) is still valid. We may then think of $\Delta w$ as
the proper thickness of the planar domain wall. Now, in the
coordinate frame of the metric (\ref{nmet}), in which the wall is
stationary, the nonvanishing components of the intrinsic metric
$h_{\mu\nu}$ and extrinsic curvature $K_{\mu\nu}$ of $\Sigma_j$
take the following simple forms:
\begin{eqnarray}\label{hcompon}
h_{\mu\nu}=g_{\mu\nu},\hspace {1cm}  \mu,\nu\neq z,
\end{eqnarray}
\begin{equation}\label{Kcompon}
K_{\mu\nu}=-\frac 1 2 g_{\mu\nu,z}.
\end{equation}
In order to find the spacetime metric on both sides of $\Sigma_j$,
we first note that within the vacuum region $\cal M_{+}$ ($\cal
M_{-}$) in which $\phi= \eta(-\eta)$, the spacetime metric can be
easily determined by solving the Einstein equations (\ref{Richi})
yielding
\begin{equation}\label{vacuum metric}
A_o(z)=-k_o |z|+ C_o,
\end{equation}
where the index $o$ denotes the quantities in the vacuum region,
and $k_o$ and $C_o$ are the relevant integration constants. Using
Eqs. (\ref{hcompon}) and (\ref{Kcompon}) we write down the
junction conditions (\ref{hmn}) and (\ref{thick israel}) as
\begin{equation}\label{k}
k=k_o,
\end{equation}
\begin{equation}\label{metric constraints1}
A_i(z)|_{z=\Delta w/2}=A_o(z)|_{z=\Delta w/2},
\end{equation}
\begin{equation}\label{metric constraints2}
\frac{\partial A_i(z)}{\partial z}|_{z=\Delta w/2}=\frac{\partial
A_o(z)}{\partial z}|_{z=\Delta w/2}.
\end{equation}
We now use the solutions (\ref{ain}), (\ref{kin}) and (\ref{vacuum
metric}) due to BCG for the wall metric in the region $\cal
M$$_{0}$, where the scalar field varies according to (\ref{inside
metric3}), and for the metric in the vacuum regions, respectively.
Then the junction conditions (\ref{metric constraints1}) and
(\ref{metric constraints2}) lead to the following constraints on
the vacuum metric constants $C_o$ and $k_o$
\begin{equation}\label{c}
C_o=1+k_o\frac{\Delta w
}{2}-\frac{\epsilon}{3}\left(2\ln(\cosh\frac{\Delta}{2})+\frac 1 2
\tanh^2\frac{\Delta}{2}\right),
\end{equation}
\begin{equation}\label{ko}
k_o=\frac {\epsilon}{w}\left(\tanh\frac{\Delta}{2}-\frac 1 3
\tanh^3\frac{\Delta}{2}\right).
\end{equation}
Note that within the context of BCG work it is supposed that the
boundaries of the wall, where the scalar field takes its vacuum
values, are at infinity. In contrast, we have modelled the thick
planar wall in such a way that the wall boundaries $\Sigma_j$ are
situated at the finite proper distances $\pm\Delta\,w/2$ from the
core of the wall. Hence, we choose $\Delta$ to be sufficiently
large in order to simulate BCG solution within our thick wall
model.
\section{From the thick to thin domain walls}
We now turn our attention to the thin wall limit of our thick wall
model. First let us define the process of passing from a thick
gravitating domain wall to a thin one by letting $\epsilon$ and
$w$ go to zero while keeping their ratio $\frac {\epsilon}{w}$
fixed. This has the effect that the distance of the event horizon
to the domain wall remains finite. The continuity of the extrinsic
curvature tensor $K_{\mu\nu}$ across the thick wall boundary, say
$\Sigma_1$, located at the proper distance $z=\Delta w/2$ as a
consequence of the Darmois junction conditions (\ref{thick
israel}) yields:
\begin{equation}\label{kmunu}
K^o_{\mu\nu}|_{z=\Delta w/2}=K^i_{\mu\nu}|_{z=\Delta w/2}.
\end{equation}
Considering  $(xx)$ component of the condition (\ref{kmunu}) and
using (\ref{Kcompon}) we can evaluate the right hand side of Eq.
(\ref{kmunu}) for the BCG wall metric solution (\ref{ain}). This
yields
\begin{equation}\label{ktt}
K^o_{xx}|_{z=\Delta
w/2}=\frac{\epsilon}{w}\left(\tanh\frac{\Delta}{2}-\frac 1 3
\tanh^3\frac{\Delta}{2}\right)\left(1-\frac{\epsilon}{3}
\left(2\ln(\cosh\frac \Delta 2)+\frac 1 2 \tanh^2\frac \Delta
2\right)\right)\exp(2k_ot),
\end{equation}
where we have also used the condition (\ref{k}). Imposing the
above thin wall limit prescription, Eq. (\ref{ktt}) reduces to
\begin{equation}\label{ktt2}
K^o_{xx}|_{z=0}=\frac
{\epsilon}{w}\left(\tanh\frac{\Delta}{2}-\frac 1 3
\tanh^3\frac{\Delta}{2}\right)\exp(2k_ot),
\end{equation}
To identify the right hand side of the equation (\ref{ktt2}) we
recall the definition of the surface energy density $\sigma$ of an
infinitely thin wall. Within our thick wall model it takes the
form
\begin{eqnarray}\label{sigmap}
\sigma=\lim_{(w\rightarrow 0, \epsilon \rightarrow 0)}\int^{\Delta
w/2}_{-\Delta w/2}\rho dz,
\end{eqnarray}
where $\rho=\rho(z)$ is the energy density of the scalar field
determined by the BCG scalar field solution (\ref{inside
metric3}). Using Eq. (\ref{sigmap}) we obtained the following
expression for $\sigma$:
\begin{eqnarray}\label{sigmap2}
\sigma=\frac{1}{2\pi
G}\frac{\epsilon}{w}\left(\tanh\frac{\Delta}{2} -\frac 1
3\tanh^3\frac{\Delta}{2}\right),
\end{eqnarray}
where we used the definition of $\epsilon$ given by
(\ref{epsilon}). Comparing the results (\ref{ktt2}) and
(\ref{sigmap2}) one immediately obtains
\begin{eqnarray}\label{thin}
K^o_{xx}|_{z=0}=2\pi G\sigma\exp(2k_ot).
\end{eqnarray}
Now, this is just the $(xx)$ component of the Israel junction
condition $K^o_{\mu\nu}=2\pi G\sigma h_{\mu\nu}$ for a planer thin
domain wall placed at $z=0$. The Israel thin wall approximation
treats the wall as a singular hypersurface with the surface energy
$\sigma$ separating the two plane symmetric vacuum spacetimes
$\cal M_{+}$ and $\cal M_{-}$ from each other. We see, therefore,
within our thick wall formulation, in the thin wall limit the
Darmois junction conditions for the extrinsic curvature tensor at
the wall boundaries generate the well-known Israel jump condition.
In the process of passing from a thick planar domain wall to the
thin one all the information about the internal structure of the
wall is squeezed in the parameter $\sigma$ characterizing the wall
surface energy density according to (\ref{sigmap2}).
\section{Evolution of the thick planar domain wall}
In the coordinate system (\ref{nmet}) used to describe the thick
wall, the wall is locally static and its core is situated at the
fixed position $z=0$. To study the evolution of the thick wall we
will look for a reference frame in which the wall is moving. In
order to do this, let us introduce a coordinate system $(v,r,X,Y)$
in which the plane symmetric metric (\ref{nmet}) with the
solutions (\ref{ain}) and (\ref{kin}) for $A$ and $k$,
respectively is transformed to the form \be\label{transin}
ds^2=f(z,t)dv^2 + 2H(z,t)dv dr-r^2(dX^2+dY^2), \ee by applying the
following appropriate coordinate transformations: \ba\label{trans}
X&=&kx,\nonumber\\
Y&=&ky,\nonumber\\
r&=&\frac{A(z)}{k}e^{kt},\nonumber\\
v&=&v(z,t), \ea where $v(z,t)$ is an unknown function to be
determined. Identification of the metrics (\ref{nmet}) and
(\ref{transin}) leads to
 \ba f(z,t) \left(\frac{\partial
v}{\partial t}\right)^2 &+&2H(z,t) \frac{\partial v}{\partial
t}\frac{\partial r}{\partial
t}=A^2(z),\nonumber\\
f( z,t) \left(\frac{\partial v}{\partial z}\right)^2 &+&2H(z,t)
\frac{\partial v}{\partial z}\frac{\partial r}{\partial
z}=-1,\nonumber\\
f( z,t) \frac{\partial v}{\partial t} \frac{\partial v}{\partial
z} &+&2H(z,t) \left(\frac{\partial v}{\partial t}\frac{\partial
r}{\partial z}+ \frac{\partial v}{\partial z}\frac{\partial
r}{\partial t}\right)=0. \ea Solving the above equations we find
\be\label{fsolve} f(z,t)=\frac{H^2\left(A^2\left(\frac{\partial
r}{\partial z}\right)^2-\left(\frac{\partial r}{\partial
t}\right)^2\right)}{A^2(z)}, \ee and \ba\label{solut}
\frac{\partial v}{\partial
t}&=&\frac{A^2}{H}\left(\frac{s_{1}A\frac{\partial r}{\partial
z}-\frac{\partial r}{\partial t}}{A^2(\frac{\partial r}{\partial
z})^2
-(\frac{\partial r}{\partial t})^2}\right),\nonumber\\
\frac{\partial v}{\partial
z}&=&-\frac{A}{H}\left(\frac{A\frac{\partial r}{\partial
z}-s_{2}\frac{\partial r}{\partial t}}{A^2(\frac{\partial
r}{\partial z})^2-(\frac{\partial r}{\partial t})^2}\right), \ea
where the sign parameters $s_{1}$ and $s_{2}$ can be taken
independently to be $\pm{1}$. Focusing our attention on the $z>0$
side of the wall, we infer that the continuity of the extrinsic
curvature on the wall boundary placed at $z=\Delta\,w/2$ ,
however, requires $s_{1}=-1$ and $s_{2}=-1$.\\
In addition, for the coordinate transformations to be integrable
we require: \be\label{integ} \frac {\partial^2 v}{\partial t
\partial z}=\frac {\partial^2 v}{\partial z\partial t}. \ee
Imposing the condition (\ref{integ}) on the solutions
(\ref{solut}) we arrive at the following: \be\label{H}
\frac{1}{H(z,t)}\frac{\partial H(z,t)}{\partial
t}-\frac{A(z)}{H(z,t)}\frac{\partial H(z,t)}{\partial z}=g(z),\ee
\be g(z)\equiv -A'(z)-k+\frac{A(z)A''(z)}{A'(z)-k}.\ee We have now
all the prerequisites to find the unknown functions defining the
transformation and to understand the dynamics of the thick wall.
Assuming a factorizable solution in the form $
H(z,t)=h_z(z)h_t(t)$, then Eq. ({\ref{H}}) leads to the two
following equations: \ba \label{ht}\frac{1}{h_t(t)}\frac{\partial
h_t(t)}{\partial t}&=&C,\\
\label{hz0}\frac{A(z)}{h_z(z)}\frac{\partial h_z(z)}{\partial
z}+g(z)&=&C, \ea where $C$ is an arbitrary constant. Integrating
Eqs. (\ref{ht}) and (\ref{hz0}) over $t$ and $z$, respectively, we
find \ba\label{Hz} H(z,t)=Dh_z(z)\exp(Ct), \ea with \ba
h_z(z)\label{hz}= \frac{2e^{(C+k)z- \frac \epsilon 6 \tanh^2\frac
z w}}{(2-\tanh\frac z w)(1+\tanh\frac z w)^2(\cosh\frac z
w)^{\frac{2\epsilon}{3}}}, \ea where $D$ is an integration
constant. The transformation (\ref{trans}) is now fully
determined, except for the explicit functional dependence of $v$
which will be considered later. The wall in the new coordinates is
defined by $R(\tau)=r(t(\tau),z=const)$ with $\tau$ being its
proper time as measured by observers at rest on the $z=const$
surface. It is then easily seen that dynamics of the thick wall in
the new coordinates is given by \be\label{eqm}
R(\tau)=\frac{A(z)}{k}e^{\frac{k\tau}{A(z)}}\Bigl|_{z=const}. \ee
This dynamical equation takes a simple form for the core of the
wall defined by $R(\tau_{0})=r(t(\tau_{0}),z=0)$. In this case we
have \be\label{eqz0} R(\tau_{0})=\frac{1}{k}e^{k\tau_{0}}. \ee
Comparing Eq. (\ref{eqz0}) with the thin wall dynamics of
\cite{ipser}, it is obvious that the core of our thick wall
evolves as if it were a thin wall with the effective surface
tension $\tilde{\sigma}=\frac{k}{2\pi G}$. In addition, taking
into account Eqs. (\ref{k}), (\ref{ko}), and (\ref{sigmap2}) it is
seen that in the thin wall limit Eq. (\ref{eqz0}) is reduced to
the Ipser-Sikivie solution for the evolution of a thin planar wall
\cite{ipser}.\\% Let us now look at the thin wall limit of the
We now intend to express the equation of motion (\ref{eqz0}) as a
function of the new coordinate $v$. Note that from the 3-metric
$h_{ab}$ induced on each $z=const$ surface of the wall we have:
\be\label{i} f(z,\tau)\dot{v}^2(z,\tau)+2\dot{R}(\tau)H(z,\tau)
\dot{v}(z,\tau)\Bigl|_{z=const}=1,\ee where dot denotes the
derivative with respect to $\tau$. Solving this for $\dot{v}$
gives \be\label{v0}
\dot{v}=\frac{-\dot{R}H+\sqrt{\dot{R}^2H^2+f}}{f}\Bigl|_{z=const},\ee
where we have made the sign choice to make sure of having a finite
solution on each $z=const$ surface specifically on the wall
boundaries where the continuity of the extrinsic curvature
requires $f$ given by the expression (\ref{fsolve}) to be zero. To
proceed further, we focus our attention on the core of the thick
wall and first write down explicitly the solution (\ref{v0}) there
\ba\label{v02} \dot{v}=\frac{e^{-(C+k)\tau_{0}}}{D }, \ea where we
have used Eqs. (\ref{fsolve}), (\ref{Hz}), (\ref{hz}), and
(\ref{eqz0}). Now, requiring that $\tau_{0}$ and $v$ point to the
same direction, i.e. $\dot{v}>0$, leads to the constraint $D>0$.
Integrating this equation over $\tau_{0}$ we get the following
results \be\label{vtaub} v(\tau_{0})=\frac{1}{D(C+k)
}\left(1-e^{-(C+k)\tau_{0}}\right),\hspace{1cm} C>-k,\ee
\be\label{vtaus} v(\tau_{0})=\frac{1}{D|C+k|
}\left(e^{|C+k|\tau_{0}}-1\right),\hspace{1cm} C<-k,\ee where we
have put $v=0$ at $\tau_{0}=0$. Inserting the solutions
(\ref{vtaub}) and (\ref{vtaus}) into Eq. (\ref{eqz0}) yields the
functional dependence of the equation of motion of the wall's core
on the time coordinate $v$ as follows: \be\label{Rvb}
R(v)=\frac{1}{k}\left(1-D(C+k)v
\right)^{\frac{-k}{C+k}},\hspace{1cm} C>-k,\ee \be\label{Rvs}
R(v)=\frac{1}{k}\left(1+D|C+k|v
\right)^{\frac{k}{|C+k|}},\hspace{1cm} C<-k.\ee It follows that
the core of the wall expands when looked upon from
the bulk frame $v$ regardless of whether $C>k$ or $C<k$.\\
To learn more about the evolution of the wall, we write down the
derivatives of Eqs. (\ref{Rvb}) and (\ref{Rvs}) with respect to
the coordinate time $v$ \ba\label{dRvb}
\frac{dR}{dv}&=&D\left(\frac{1}{1-D(C+k)v
}\right)^{\frac{2k+C}{C+k}},
\hspace{1.5cm}  C>-k,\\
\label{dRvc}\frac{dR}{dv}&=&D\left(
1+D|C+k|v\right)^{\frac{k}{|C+k|}-1},\hspace{1cm} C<-k.\ea
Consequently, from  Eqs. (\ref{dRvb}) and (\ref{dRvc}) we see that
the expansion of the wall's core slows down provided $C<-2k$,
otherwise it speeds up.\\
From the coordinate transformation we have chosen to investigate
the dynamics of the wall, it is clear that the wall thickness is
no longer a constant . To see the time dependence of the thickness
we first note that the spacetime parts $(z>0)$ and $(z<0)$ are
transformed separately to spheres. Therefore, it is more
appropriate to look at the half-thickness ($w^*$) defined by
\be\label{w*} w^*=R(v)\left|_{z=0}-R(v)\right|_{z=\frac{\Delta
w}{2}}.\ee Following the above procedure leading to the equations
of motion for the core of the wall one can find the corresponding
equation for the boundary of the wall located at $z=\frac{\Delta
w}{2}$. We get \be\label{Rvbz} R(v)|_{z=\frac{\Delta
w}{2}}=\frac{A}{k}\left(1-\frac{2D(C+k)h_z(z)}{A}(v-v_0)
\right)^{\frac{-k}{C+k}}\big|_{z=\frac{\Delta w}{2}},\ee where we
have considered the case $C>-k$, $v_{0}$ denotes the initial value
of $v$ when the proper time on the boundary of the wall is taken
to be zero. Plugging Eqs. (\ref{Rvb}) and (\ref{Rvbz}) into the
definition (\ref{w*}) we end up with  \ba
w^*&=&\frac{1}{k}\left(1-D(C+k)v
\right)^{\frac{-k}{C+k}}\nonumber\\
&-&\frac{A}{k}\left(1-\frac{2D(C+k)h_z(z)}{A}(v-v_0)
\right)^{\frac{-k}{C+k}}\big |_{z=\frac{\Delta w}{2}}, \ea where
$A(z)$ and $h_z(z)$ are given by the solutions (\ref{ain}) and
(\ref{hz}), respectively. This gives the time dependence of the
half-width of the wall in terms of the new coordinate $v$. Noting
that $\frac{\partial r}{\partial z}<0$, we can see that the
half-width $w^*$ is an increasing function of $v$.
\section{Conclusion}
We have studied the thick planar domain wall as a spacetime having
two boundaries at the same proper distance from the wall's core,
as first formulated in \cite{khak02}. It has then been shown that
in the thin wall limit the Darmois junction conditions generate
the familiar Israel's jump condition at the boundary of the
corresponding thin wall with the embedding spacetimes. It is
realized that in the limiting process to the thick  planar domain
wall all the information about the internal structure of the wall
is squeezed in the surface energy density $\sigma$ introduced in
(\ref{sigmap2}). We have also given explicitly a coordinate
transformation from the plane symmetric metric (\ref{nmet})
describing a thick planar domain wall at rest to a reference frame
in which the thick wall evolves. Starting from an ansatz for the
transformed metric of the form (\ref{transin}) this aim was
achieved by analytically solving the transformation equations. It
is then seen that the core of the thick wall is moving with
respect to the transformed reference frame as if it were a thin
wall. In contrast, according to Eq. (\ref{eqm}) the other layers
of the wall evolve differently. We have also studied the thin wall
limit of the moving thick wall and showed that in this limit the
thick wall's solution becomes the well-known thin wall solution of
\cite{ipser}. In this dynamical picture, the wall thickness shows
a time variation, which has been given explicitly.

\end{document}